\documentclass[aps,amsmath,amssymb,amsfonts,showpacs,twocolumn,pra]{revtex4}
\usepackage{amscd}
\usepackage{amsthm}
\usepackage{graphicx}

\def\qed{\leavevmode\unskip\penalty9999 \hbox{}\nobreak\hfill
     \quad\hbox{\leavevmode  \hbox to.77778em{%
               \hfil\vrule   \vbox to.675em%
               {\hrule width.6em\vfil\hrule}\vrule\hfil}}
     \par\vskip3pt}

\def\ra{\rangle}
\def\la{\langle}
\def\be{\begin{equation}}
\def\ee{\end{equation}}
\def\ba{\begin{array}}
\def\ea{\end{array}}

\begin{document}
\title{Unextendible maximally entangled bases in $\mathbb{C}^{d}\bigotimes\mathbb{C}^{d}$}
\author{Yan-Ling Wang$^{1}$, Mao-Sheng Li$^{1}$, Shao-Ming Fei$^{2, 3}$}

 \affiliation
 {
   {$^1$Department of Mathematics,
 South China University of Technology, Guangzhou
510640, P.R.China} \\
{$^2$School of Mathematical Sciences, Capital Normal University,
Beijing 100048, China}\\
{$^3$Max-Planck-Institute for Mathematics in the Sciences, 04103
Leipzig, Germany}
}

\begin{abstract}

We investigate the unextendible maximally entangled bases in $\mathbb{C}^{d}\bigotimes\mathbb{C}^{d}$
and present a $30$-number UMEB construction in  $\mathbb{C}^{6}\bigotimes\mathbb{C}^{6}$.
For higher dimensional case, we show that for a given $N$-number UMEB in $\mathbb{C}^{d}\bigotimes\mathbb{C}^{d}$,
there is a $\widetilde{N}$-number, $\widetilde{N}=(qd)^2-q(d^2-N)$, UMEB in $\mathbb{C}^{qd}\bigotimes\mathbb{C}^{qd}$
for any $q\in\mathbb{N}$. As an example, for $\mathbb{C}^{12n}\bigotimes\mathbb{C}^{12n}$ systems,
we show that there are at least two sets of UMEBs which are not equivalent.
\end{abstract}

\maketitle
\maketitle

\section{Introduction}
Einstein, Podolsky, and Rosen (EPR) proposed a thought experiment which demonstrated that quantum mechanics is not a complete theory of nature \cite{EPR,nils},
quantum entanglement has been shown to be tightly related to some fundamental problems in quantum mechanics such as reality and nonlocality.
It was quite surprising when it was found that there are sets of product states which nevertheless display a form of nonlocality \cite{BD1,BD2}.
It was shown that there are sets of orthogonal product vectors  in {\small $\mathbb{C}^{m}\bigotimes\mathbb{C}^{n}$}
such that there are no further product states which are orthogonal to all the state in the set, even though the space spanned by the set
is smaller than $nm$. A set of states satisfying such property is called unextendible product bases (UPBs).
Many useful applications have been obtained ever since the concept of UPBs in multipartite quantum systems was introduced \cite{mhor, EofCon,Be}.
It was shown that the UPBs are not distinguishable by local measurements and classical communication, and the space complementary
to a UPB contains bound entanglement \cite{Be}.

In 2009, S. Bravyi and J. A. Smolin generalized the notion of the UPB to unextendible maximally entangled basis \cite {s3}: a set of orthonormal
maximally entangled states in {\small$\mathbb{C}^{d}\bigotimes\mathbb{C}^{d}$} consisting of fewer than $d^2$ vectors which have no
additional maximally entangled vectors that are orthogonal to all of them. The authors proved that there do not exist UMEBs
for $d=2$, and constructed a 6-member UMEB for $d=3$ and a 12-member UMEB for $d=4$.

In Ref. \cite{BC}, B. Chen and S.M. Fei studied the  UMEB in {\small $\mathbb{C}^{d}\bigotimes\mathbb{C}^{d'}$} ($\frac{d'}{2}<d<d'$).
They constructed a $d^{2}$-member UMEBs, and left an open problem for the existence of UMEBs in the case of $\frac{d'}{2}\geq d$ .
Recently, we  give an explicit construction of UMEB in
{\small$\mathbb{C}^{d}\bigotimes\mathbb{C}^{d'} (d<d')$} \cite{Li}. We show that the states in the complementary
space of the UMEBs have Schmidt numbers less than $d$.

In this paper, we study the unsolved problem of UMEBs in $\mathbb{C}^{d}\bigotimes\mathbb{C}^{d}$.
We start with the construction of a 30-member UMEB in {$\mathbb{C}^{6}\bigotimes\mathbb{C}^{6}$.
Then we generalized the example to higher dimension case.
We show that for an given $N$-number UMEB in $\mathbb{C}^{d}\bigotimes\mathbb{C}^{d}$},
there is a $\widetilde{N}$-number, $\widetilde{N}=(qd)^2-q(d^2-N)$, UMEB in {$\mathbb{C}^{qd}\bigotimes\mathbb{C}^{qd}$}
for any $q\in\mathbb{N}$. For {$\mathbb{C}^{12n}\bigotimes\mathbb{C}^{12n}$} systems,
we show that there are at least two sets of UMEBs which are not equivalent.

\section{UMEBs in $\mathbb{C}^{d}\bigotimes\mathbb{C}^{d}$}

A set of states \{$|\phi_{a}\rangle\in\mathbb{C}^{d}\bigotimes\mathbb{C}^{d}:\,a=1,2,\cdots,n,\,n<d^2$\}
is called an $n$-number UMEB if and only if
(i) $|\phi_{a}\rangle$, $a=1,2,\cdots,n$, are maximally entangled;
(ii) $\langle\phi_{a}|\phi_{b}\rangle=\delta_{ab}$;
(iii) if $\langle\phi_{a}|\psi\rangle=0$ for all $a=1,2,\cdots,n$, then $|\psi\rangle$ cannot be maximally entangled.

Here under computational basis a maximally entangled state $|\phi_{a}\rangle$ can be expressed as
\be\label{1}
|\phi_{a}\rangle=(I\otimes U_a)\,\frac{1}{\sqrt{d}}\sum_{i=1}^d|i\ra\otimes |i\ra,
\ee
where $I$ is the $d\times d$ identity matrix, $U_a$ is any unitary matrix. According to (\ref{1}),
a set of unitary matrices {$\{U_a\in M_d(\mathbb{C})|a=1,...,n\}$} gives an $n$-number UMEB in {$\mathbb{C}^{d}\bigotimes\mathbb{C}^{d}$} if and only if\\
 (i)  $n<d^2$;\\
 (ii) {$Tr(U_a^\dag U_b)=d\,\delta_{ab},~~ \forall  a,b=1,\cdots,n$;\\}
 (iii) For any {$U\in M_d(\mathbb{C}),$} if {$ Tr(U_a^\dag U)=0,~\forall\,  a=1,\cdots,n$,} then {$U$} cannot be unitary.\\

In the following we present a 30-member UMEB in {$\mathbb{C}^{6}\bigotimes\mathbb{C}^{6}$}. Set
$$U_{nm}\triangleq \sum_{k=0}^2 e^{\frac{2\pi\sqrt{-1}}{3}kn}|k\oplus m \rangle \langle k |,$$
$$
U_{nm}^{\pm }=\delta_{\pm} \otimes U_{nm} \ \  n,m=1,2,3,
$$
and
$$
U_{i}^{\pm }=\eta_{\pm} \otimes U_{i} \ \  i=1,2,3,4,5,6,
 $$
where $k\oplus m$ denotes the number $k+m$ mod $d$,
$$
\delta_{\pm}=\left(
                 \begin{array}{cc}
                   0 & 1 \\
                   \pm 1 & 0 \\
                 \end{array}
               \right),~~~~
\eta_{\pm}=\left(
                 \begin{array}{cc}
                   1 & 0 \\
                   0 & \pm 1 \\
                 \end{array}
               \right),
$$
{$\{U_i\}_{i=1}^{6}$ } are the unitary matrices constructed in {$\mathbb{C}^{3}\bigotimes\mathbb{C}^{3} $} Ref. \cite{s3}:
$$
U_i=I-(1-e^{i\theta})|\psi_i\ra\la\psi_i|,~~~i=1,2,...,6,
$$
where
$$
\ba{l}
|\psi_{1,2}\ra=\displaystyle\frac{1}{\sqrt{1+\alpha^2}}(|0\ra\pm|1\ra),\\[3mm]
|\psi_{3,4}\ra=\displaystyle\frac{1}{\sqrt{1+\alpha^2}}(|1\ra\pm|2\ra),\\[3mm]
|\psi_{5,6}\ra=\displaystyle\frac{1}{\sqrt{1+\alpha^2}}(|2\ra\pm|0\ra),
\ea
$$
with $\alpha=(1+\sqrt{5})/2$.

We now prove that $\{U_{nm}^{\pm }$, $U_{i}^{\pm }$, $n, m=1,2,3$; $i=1,...,6 \}$ give rise to a 30-member UMEB in {$\mathbb{C}^{6}\bigotimes\mathbb{C}^{6}$}.

(1) Since $\{U_{nm}\}$ and $\{U_i\}$ are  unitary, it is easily seen that $\{U_{nm}^{\pm } , U_{i}^{\pm }\}$ are also unitary.
(2) To prove the orthogonality of these unitary states, we consider three different cases:\\
(i) inner product between two elements in $\{U_{nm}^{\pm }\}:$
$$
\ba{rcl}
Tr({(\delta_+ \otimes U_{nm})}^\dag (\delta_{\pm} \otimes U_{\widetilde{n} \widetilde{m}}))
&=&\pm\,Tr(\eta_{\pm} \otimes U_{nm}^{\dag}U_{\widetilde{n}\widetilde{m}})\\[2mm]
&=&6 \delta_{+\pm} \delta_{n\widetilde{n}}\delta_{m\widetilde{m}};
\ea
$$
(ii) inner product between two elements in  $\{U_{i}^{\pm }\}:$
$$Tr({(\eta_+ \otimes U_{i})}^\dag
               \eta_{\pm} \otimes U_{\widetilde{i} })
=Tr(\eta_{+}\eta_{\pm} )
Tr(U_{i}^{\dag}U_{\widetilde{i}})
=6 \delta_{+\pm} \delta_{i\widetilde{i}};$$
(iii) the inner product between one elements in $\{U_{nm}^{\pm }\}$ and the one in $\{U_{i}^{\pm }\}:$
$$Tr({(\delta_{pm} \otimes U_{nm})}^\dag
               \eta_{\pm} \otimes U_{i})
=Tr(\delta_{\pm}\eta_{\pm})Tr(U_{nm}^{\dag}U_{i})=0.$$
(3) Assume that {\small $U\in M_6(\mathbb{C})$ }satisfy:
$$
Tr(U^\dag U_{nm}^\pm)=0~ \text{ and }~ Tr(U^\dag U_{i}^\pm)=0.
$$
Let $V_1=\text{span}\{U_{nm}^\pm\}$, dim$V_1=18$. Denote
$$
V_2=\left\{\left[
              \begin{array}{cc}
                A & \bf{0} \\
                \bf{0} & B \\
              \end{array}
            \right]|A,B\in M_3(\mathbb{C})\right\},
$$
then $\text{dim}V_2=18$. Since the canonical inner product
 $$
 Tr\left({\left[
       \begin{array}{cc}
         A & \bf{ 0} \\
         \bf{ 0 }& B \\
       \end{array}
     \right]}^\dag
  \left[
       \begin{array}{cc}
         \bf{0} & U_{nm} \\
         \pm U_{nm} & \bf{0} \\
       \end{array}
     \right]\right)=0,
 $$
one has $V_1^{\bot}=V_2$. Now let $V_3=\text{span}\{U_{nm}^\pm,U_{i}^\pm\}$.
We have dim$V_3=30$ and $V_3^{\bot}\subset V_1^{\bot}=V_2$.
Therefore $U\in V_3^{\bot}$, and the matrix $U$ has the form
$U=diag(W_1, W_2)$, where $W_1, W_2 \in M_3(\mathbb{C})$.
As $U$ satisfies
$$
Tr\left({\left[
       \begin{array}{cc}
         W_1 & \bf 0 \\
         \bf 0 & W_2 \\
       \end{array}
     \right]}^\dag
  \left[
       \begin{array}{cc}
         U_i & \bf 0 \\
         \bf 0 & \pm U_i \\
       \end{array}
     \right]\right)=0,
$$
i.e. $Tr({W_1}^\dag U_i)\pm Tr({W_2}^\dag U_i)=0$,
we have $Tr({W_1}^\dag U_i)=Tr({W_2}^\dag U_i)=0$ for $i=1,2,\cdots,6$, which implies that $W_1,W_2\notin U(3)$. Hence
$U\notin U(6)$. Therefore we conclude that $\{U_{nm}^\pm,U_{i}^\pm\}$ is a 30-member UMEB in {$\mathbb{C}^{6}\bigotimes\mathbb{C}^{6}$}.  \qed

Now we show that for any UMEB in {\small$\mathbb{C}^{d}\bigotimes\mathbb{C}^{d}$}, there will be an UMEB
in {\small$\mathbb{C}^{qd}\bigotimes\mathbb{C}^{qd}$} for any $q\in\mathbb{N}$.

\noindent{\bf Theorem 1.} If there is an $N$-number UMEB in {\small$\mathbb{C}^{d}\bigotimes\mathbb{C}^{d}$}, then for any $q\in\mathbb{N}$,
there is a $\widetilde{N}$-number, $\widetilde{N}=(qd)^2-q(d^2-N)$, UMEB in {\small$\mathbb{C}^{qd}\bigotimes\mathbb{C}^{qd}.$}

\noindent \emph{Proof:} Denote

 {\footnotesize
 $$ S=\left[
                 \begin{array}{ccccc}
                   0 & 1 & 0 & \cdots & 0 \\
                   0 & 0 & 1 & \cdots & 0 \\
                   \vdots & \vdots & \vdots & \ddots & \vdots \\
                   0 & 0 & 0 & \cdots & 1 \\
                   1 & 0 & 0 & \cdots & 0 \\
                 \end{array}
               \right],
                {W}=\left[
                 \begin{array}{ccccc}
                   1 & 1 & 1 & \cdots & 1 \\
                   1 & \zeta_q & \zeta_q^2 & \cdots & \zeta_q^{q-1} \\
                   1 & \zeta_q^2 & \zeta_q^4 & \cdots  & \zeta_q^{2(q-1)} \\
                   \vdots & \vdots & \vdots & \ddots & \vdots \\
                   1 & \zeta_q^{q-1} & \zeta_q^{2(q-1)} & \cdots & \zeta_q^{(q-1)^2}\\
                 \end{array}
               \right],
 $$}

 where $\zeta_q=e^{\frac{2\pi\sqrt{-1}}{q}}$ and
 $$
 U_{nm}=\sum_{k=0}^{d-1} e^{\frac{2\pi\sqrt{-1}}{d}kn}|k\oplus m \rangle \langle k |,\ m,n=0,1,\cdots,d-1.
 $$
In the following for any $q\times q$ matrix $M$ with entries $m_{ij}$,
we define $M^i=\text{diag}(m_{i+1,1},m_{i+1,2},...,m_{i+1,q})$, $i\in\{0,1,\cdots,q-1\}$. In order to simplify notation, we suppose {\small $W=\{w_{ij}\}_{i,j=1}^p$}.

Let $\{U_n\}$, $n=1,2,\cdots,N<d^2$, be the set of unitary matrices that give rise to the UMEB in {\small$\mathbb{C}^{d}\bigotimes\mathbb{C}^{d}$}.
Set
$$
U_{nm}^{ij}=(W^i S^j)\otimes U_{nm},
$$
where $\ 0\leq i \leq q-1, 1\leq j \leq q-1,m,n=0,\cdots,d-1 $, and
$$
U_n^i=W^i\otimes U_n,i=0,1,\cdots,q-1,n=1,2,\cdots,N<d^2.
$$
Let $\widetilde{N}$ denote the number of matrices in
{\small$\{U_{nm}^{ij},U_{n}^{i}\}$}. We have
$$
\widetilde{N}=q(q-1)d^2+qN=(qd)^2-q(d^2-N)<q^2d^2.
$$

Next we prove that {\small$\{U_{nm}^{ij},U_{n}^{i}\}$} give a $\widetilde{N}\text{-member}$ UMEB in {\small$\mathbb{C}^{qd}\bigotimes\mathbb{C}^{qd}$}.

(1) Since {\small$W^i,S^j,U_{nm}$} are all unitary, so are {\small$\{U_{nm}^{ij},U_{n}^{i}\}$}. So the given set of matrices satisfy the first condition of UMEB.

(2) In order to  prove the orthogonality of the related basic states, we need to check the inner products between two elements in {\small$\{U_{nm}^{ij}\}$},
between two elements in {\small$\{U_{n}^{i }\}$}, and between one in {\small$\{U_{nm}^{ij }\}$} and the other one in {\small$\{U_{n}^{i }\}$}. It is direct to verify that\\
(i)  $Tr({(U_{nm}^{ij})}^\dag U_{\widetilde{n}\widetilde{m}}^{\widetilde{i}\widetilde{j}})
=qd\delta_{i\widetilde{i}}\delta_{j\widetilde{j}}\delta_{n\widetilde{n}}\delta_{m\widetilde{m}}$;\\
(ii) $Tr({(W^i\otimes U_n)}^\dag (W^{\widetilde{i}}\otimes U_{\widetilde{n}}))=qd\delta_{i\widetilde{i}}\delta_{n\widetilde{n}}$;\\
(iii)  $Tr({(U_{nm}^{ij})}^\dag U_{\widetilde{n}}^{\widetilde{i}})=Tr({(S^j)}^\dag{(W^i)}^\dag W^{\widetilde{i}}\otimes U_{nm}^\dag U_{n})=0$.

(3) Let {\small$V_1=\text{span}\{U_{nm}^{ij}\}$} be a subspace of {\small$M_{qd}(\mathbb{C})$}, $\text{dim}V_1=q(q-1)d^2$.
Denote
$$
V_2=\left\{diag(A_1,A_2,...,A_q)| A_i\in M_d(\mathbb{C}),~i=1,2,...,q\right\}.
$$
It is seens that dim$V_2=qd^2.$ For any matrix {\small$A\in V_2$ }and $\ 0\leq i \leq q-1, 1\leq j \leq q-1,m,n=0,\cdots,d-1, $
we have {\small$ Tr(A^\dag U_{nm}^{ij})=0.$}
Thus for any matrix {\small$A\in V_2$} and {\small$B\in V_1$}, {\small$Tr(A^\dag B)=0.$}  Namely, {\small$V_2\subseteq V_1^{\bot} $}.
Accounting to the dimensions of {\small$ V_1,V_2 \text{ and } M_{qd}(\mathbb{C})$}, we obtain  {\small$V_1^{\bot}=V_2.$}
Set {$V_3=\text{span}\{U_{nm}^{ij},U_n^i\}.$}
Clearly,{\small $ V_3^{\bot}\subset V_1^{\bot} $}. Hence any {\small$U\in V_3^{\bot}$ }has the following form
$$
U=diag\left(W_1,W_2,...,W_q\right) \ \ \text{ where } W_i\in M_d(\mathbb{C}).
$$

In addition, from {\small$Tr(U^{\dag}U_n^i)=0$}, for $i=1,...,q$, we have
{\small
 $$Tr(\left[
            \begin{array}{ccc}
              W_1 &  &  \\
                & \ddots &  \\
                &  & W_q \\
            \end{array}
          \right]
          \left[
            \begin{array}{ccc}
              w_{i1}U_n &  &  \\
                & \ddots &  \\
                &  & w_{iq}U_n \\
            \end{array}
          \right])=0.$$
          }
i.e.,
{\small\begin{equation}{\label{Equa:2}}
w_{i1}Tr(W_1^{\dag}U_n)+\cdots+w_{iq}Tr(W_q^{\dag}U_n)=0,\
i=1,...,q.
\end{equation}
}
Noting that
 {\small$$\det(W)=\det
             \left[
                 \begin{array}{ccccc}
                   1 & 1 & 1 & \cdots & 1 \\
                   1 & \zeta_q & \zeta_q^2 & \cdots & \zeta_q^{q-1} \\
                   1 & \zeta_q^2 & \zeta_q^4 & \cdots  & \zeta_q^{2(q-1)} \\
                   \vdots & \vdots & \vdots & \ddots & \vdots \\
                   1 & \zeta_q^{q-1} & \zeta_q^{2(q-1)} & \cdots & \zeta_q^{(q-1)^2}\\
                 \end{array}
               \right]
\neq 0,$$}
from equation (\ref{Equa:2}) we obtain
{\small $$ Tr(W_1^{\dag}U_n)=\cdots=Tr(W_q^{\dag}U_n)=0,  \text{ for } n=1,\cdots,N.$$}
Therefore {\small{$W_i\notin U(d)$}}, and hence {\small $U\notin U(qd)$.}
From (1), (2) and (3), we conclude that {\small$\{U_{nm}^{ij},U_n^i\}$} is an {$\widetilde{N}\text{-member}$} UMEB in
{\small $\mathbb{C}^{qd}\bigotimes\mathbb{C}^{qd}.$ } \qed

\noindent{{\bf Corollary 1.}} In {\small$\mathbb{C}^{3n}\bigotimes\mathbb{C}^{3n}$}, there exists an UMEB.\\
\noindent{{\bf Corollary 2.}} In {\small$\mathbb{C}^{4n}\bigotimes\mathbb{C}^{4n}$}, there exists an UMEB.\\

In \cite {s3} a 6-member UMEB for $d=3$ and a 12-member UMEB for $d=4$ have been constructed.
We have constructed in this paper a 30-member UMEB for $d=6$.
From our theorem, for $d=12$, one con construct $\widetilde{N}=(qd)^2-q(d^2-N)$-number UMEBs in {\small $\mathbb{C}^{qd}\bigotimes\mathbb{C}^{qd}$},
by respectively taking $N=3,4,6$ and $q=4,3,2$. Therefore in {$\mathbb{C}^{12}\bigotimes\mathbb{C}^{12}$} there are three
ways to construct UMEBs from the UMEBs of dimension 3,4 and 6. In the following we show that at least two of the three UMEBs obtained in this way are not equivalent.

\noindent{\bf Definition 1} Let {\small$\{|\psi_a\rangle\}_{a=1}^n$} and {\small$\{|\phi_a\rangle\}_{a=1}^n$ } be two sets of  UMEBs in {\small$\mathbb{C}^{d}\bigotimes\mathbb{C}^{d}$}.   They are called  equivalent if
 {\small$\exists $ $ \sigma \in S_n$, $U,V\in U(n)$ } such that
 {$U\otimes V |\psi_a\rangle  =|\phi_{\sigma(a)}\rangle$ for $a=1,..,n.$}
 (Here $S_n$ is the permutation group of $n$ elements)

Two sets of UMEBs are equivalent means
 {\small$\exists $ $ \sigma \in S_n$, $U,V\in U(n)$ } such that
 {\small$U\otimes V |\psi_a\rangle  =|\phi_{\sigma(a)}\rangle$ for $a=1,..,n.$}
  That is, {\small$(U\otimes V) (I\otimes U_a) |\psi\rangle  =I\otimes V_{\sigma(a)} |\psi\rangle$}, or equivalent, {\small$U\otimes V_{\sigma(a)}^\dag V  U_a |\psi\rangle=|\psi\rangle$}. But the invariant group of the state  $|\psi\rangle$ is the form {\small$U\otimes U^*$}, where {\small$U^*$} means the conjugate of {\small$U$}. So we can give an equivalent definition.

 \noindent{\bf Definition $1'$} Two sets of  UMEBs {\small$\{U_a\}_{a=1}^n$} and {\small$\{V_a\}_{a=1}^n$ } in {\small$\mathbb{C}^{d}\bigotimes\mathbb{C}^{d}$}  are called  equivalent if {\small$\exists $ $ \sigma \in S_n$, $U,V\in U(n)$ }such that  {$U U_a V  =V_{\sigma(a)}$} for $a=1,..,n.$

We can deduce that {\small$U U_a U_b U^\dag  =V_{\sigma(a)}V_{\sigma(b)}.$} Now
 we consider the two sets of UMEBs in  {\small$\mathbb{C}^{12}\bigotimes\mathbb{C}^{12}$},
 let {\small$\{U_1,...,U_6\}$ }be the 6-member UMEB found in  {\small$\mathbb{C}^{3}\bigotimes\mathbb{C}^{3} $} Ref \cite{s3}. We notice that the eigenvalues of  {\small${U_1,...,U_6}$} are all $\{ 1, 1 , e^{\sqrt{-1}\theta}\}$ where $\cos\theta=-\frac{7}{8}$. But in Ref. \cite{tran1,tran2} we can see that $\cos^2(\frac{n}{m}\pi)\in \mathbb{Q}$ if and only if  $\cos^2(\frac{n}{m}\pi)\in \{0, \frac{1}{4},\frac{1}{2},\frac{3}{4},1\}$ (where $\mathbb{Q}$  is the set of rational numbers). So $\cos^2\theta=\frac{49}{64}$ implies that $\theta$ is not of the form $\frac{n}{m}\pi$. Hence for any $n\in \mathbb{N}, {(e^{\sqrt{-1}\theta})}^n\neq 1.$  Since  {\small$U_n^i =W^i\otimes U_n$}, the eigenvalues of  {\small$U_n^i$} are $\{ 1,...,\zeta_4^{3i},1,...,\zeta_4^{3i},e^{\sqrt{-1}\theta} ,...,e^{\sqrt{-1}\theta}\zeta_4^{3i}\}.$ If we consider the order of the eigenvalue $\lambda$ (i.e. the least nature number $n $ such that $\lambda^n=1$ ), then the orders of the four $\{\zeta_4^{3i},e^{\sqrt{-1}\theta} ,...,e^{\sqrt{-1}\theta}\zeta_4^{3i}\} $ are infinite. Then there are four eigenvalues of each {\small$U_{m,n}^{i,j}(U_{\widetilde{n}}^{\widetilde{i}})^\dag$} with order infinite. And the order of eigenvalues of {\small$U_{m,n}^{i,j}(U_{\widetilde{m},\widetilde{n}}^{\widetilde{i},\widetilde{j}})^\dag$} and {\small$U_{n}^{i}(U_{\widetilde{n}}^{\widetilde{i}})^\dag$} are all finite.  Similarly, we can calculate the orders of eigenvalues of  {\small$U_{m,n}^{i,j}(U_n^i)^\dag$},
 {\small$U_{m,n}^{i,j}(U_{\widetilde{m},\widetilde{n}}^{\widetilde{i},\widetilde{j}})^\dag$} and {\small$U_{n}^{i}(U_{\widetilde{n}}^{\widetilde{i}})^\dag$} derived from the UMEB in {\small$\mathbb{C}^{4}\bigotimes\mathbb{C}^{4}$ }. The matrices with infinite order of eigenvalues are just of the form  {\small$U_{m,n}^{i,j}(U_{\widetilde{n}}^{\widetilde{i}})^\dag$} or {\small$U_{n}^{i}(U_{\widetilde{n}}^{\widetilde{i}})^\dag$} with $\widetilde{n}=3,4,5.$ The number of matrices with infinite order of the eigenvalues are presented in Table I. Then  we can easily judge that the above two sets of UMEBs are not equivalent from the formula {\small$U U_a U_b U^\dag  =V_{\sigma(a)}V_{\sigma(b)}$}.

 {\small
\begin{table}
 \caption{ number of matrices with infinite order of eigenvalues in $\mathbb{C}^{12}\bigotimes\mathbb{C}^{12}$}
$\begin{array}{ccccc}\hline\hline
   & \text{derive from } & U_{m,n}^{i,j}(U_{\widetilde{n}}^{\widetilde{i}})^\dag & U_{m,n}^{i,j}(U_{\widetilde{m},\widetilde{n}}^{\widetilde{i},\widetilde{j}})^\dag & U_{n}^{i}(U_{\widetilde{n}}^{\widetilde{i}})^\dag \\ \hline
      &  \mathbb{C}^{3}\bigotimes\mathbb{C}^{3} &  2592 & 0 & 0 \\
      &  \mathbb{C}^{4}\bigotimes\mathbb{C}^{4} & 270 & 0 & 180 \\ \hline\hline
\end{array}$
\end{table}
}

Moreover, We notice that the above conclusion can be generalized to  {\small$\mathbb{C}^{12n}\bigotimes\mathbb{C}^{12n}$}. In the UMEB derived from that in {\small$\mathbb{C}^{3}\bigotimes\mathbb{C}^{3}$ }, there are $24n$ matrices of the form  {\small$U_{n}^{i}$} and $36n(4n-1)$ matrices of the form  {\small$U_{m,n}^{i,j}$}, then there are $864n^2(4n-1)$ elements of the form $U_a U_b$ of the UMEB  with infinite order eigenvalues. In the UMEB derived from that in {\small$\mathbb{C}^{4}\bigotimes\mathbb{C}^{4}$ }, there are $36n$ matrices of the form  {\small$U_{n}^{i}$} and $48n(3n-1)$ matrices of the form  {\small$U_{m,n}^{i,j}$}, but there are at most $405n^2(3n-1)+180n^2$ elements of the UMEB  with infinite order eigenvalues. Hence they are not equivalent. We just give a corollary without proof.\\
\noindent{{\bf Corollary 3.}} In  {\small$\mathbb{C}^{12n}\bigotimes\mathbb{C}^{12n}$}, there exist two sets of UMEBs which are not equivalent.
\section{conclusion}
We have studied the UMEBs in {$\mathbb{C}^{d}\bigotimes\mathbb{C}^{d}$} and presented a $30$-number UMEB construction in  $\mathbb{C}^{6}\bigotimes\mathbb{C}^{6}$.
By using approach in \cite{Li}, we have presented the construction of an UMEB in  {$\mathbb{C}^{qd}\bigotimes\mathbb{C}^{qd} $} from an UMEB in {$\mathbb{C}^{d}\bigotimes\mathbb{C}^{d}$}. In particular, we can obtain UMEBs in {$\mathbb{C}^{3n}\bigotimes\mathbb{C}^{3n}$ } and  {$\mathbb{C}^{4n}\bigotimes\mathbb{C}^{4n}$} from the results in \cite {s3}.
By analysing the order of the eigenvalues of UMEB in {$\mathbb{C}^{12}\bigotimes\mathbb{C}^{12}$} derived from the UMEBs in
{$\mathbb{C}^{3}\bigotimes\mathbb{C}^{3}$} and in {$\mathbb{C}^{4}\bigotimes\mathbb{C}^{4}$}, it has been shown that the two sets of UMEBs in  {$\mathbb{C}^{12}\bigotimes\mathbb{C}^{12}$}, obtained from our theorem, are not equivalent. Similarly there are two sets of UMEBs in  {$\mathbb{C}^{12n}\bigotimes\mathbb{C}^{12n}$} which are not equivalent.
As a summary, Table II shows the known results about the UMEBs in $\mathbb{C}^{d}\bigotimes\mathbb{C}^{d'}$.

 {\small
\begin{table}
 \caption{ Results about UMEBs in  {\small$\mathbb{C}^{d}\bigotimes\mathbb{C}^{d'}$}}
 $\begin{array}{ccc}\hline\hline
   \text{condition} & \text{number in UMEB} & \text{reference}   \\ \hline
   d=d'=2 & \text{none} &  [8] \\
   d=d'=3 & 6 &    [8]\\
   d=d'=4 & 12 &     [8]\\
   d<d'<2d & d^2 &     [9]\\
   d'=qd+r,0<r<d & qd^2 &   [10]  \\
   d'> d & d(d'-1) &      [10]\\
   d=d'=3n & d(d-1) &    \text{This paper} \\
   d=d'=4n & d(d-1) &    \text{This paper}\\ \hline\hline
 \end{array}$
\end{table}
}

\vspace{2.5ex}
\noindent{\bf Acknowledgments}\, \,
This work is supported by the NSFC under number 11275131.

\end{document}